\documentclass[aps,prb,onecolumn,notitlepage,10pt,superscriptaddress,nofootinbib]{revtex4-2}

\usepackage[letterpaper,margin=1in]{geometry}
\usepackage{amsmath,amssymb,amsfonts,bm,mathtools}
\usepackage{graphicx}
\usepackage{booktabs}
\usepackage{xcolor}
\usepackage{microtype}
\usepackage{tikz}
\usetikzlibrary{arrows.meta,positioning,calc}
\usepackage[colorlinks=true,linkcolor=blue,citecolor=blue,urlcolor=blue]{hyperref}

\graphicspath{{figures/}}

\newcommand{\qv}{\mathbf q}

\newcommand{\rr}{\mathbf r}
\newcommand{\RR}{\mathbf R}
\newcommand{\ee}{\mathrm e}
\newcommand{\ii}{\mathrm i}
\newcommand{\dd}{\mathrm d}
\newcommand{\order}{\mathcal O}
\newcommand{\intq}{\int\!\frac{\dd^2q}{(2\pi)^2}}
\newcommand{\eps}{\varepsilon}
\newcommand{\calT}{\mathcal T}
\newcommand{\calL}{\mathcal L}
\newcommand{\calN}{\mathcal N}

\hypersetup{
  pdftitle={Thermal and viscous contrast in quantum Hall scanning-probe images},
  pdfauthor={P. Shubham Parashar}
}

\begin{document}

\title{Thermal and viscous contrast in quantum Hall scanning-probe images}
\author{P.~Shubham~Parashar}
\email{psparash@ucsd.edu}
\email{pathaksparash@gmail.com}
\affiliation{Independent Researcher, San Diego, California, USA}
\date{July 2026}

\begin{abstract}
Quantum Hall scanning images are often read as maps of a local potential, temperature, or viscosity, whereas a probe in fact records a finite-resolution functional of a transport operator.  We formulate that functional explicitly, combining Landau-level projection, a particle-number Ward identity, magnetization-subtracted thermoelectric transport, a hydrodynamic Stokes--Ohm inversion, and finite-tip Fisher information.  Two results follow in complementary transport regimes.  In the strong-field, sharp-Landau-level regime, the defect-induced thermoelectric and electrical Hall contrasts of a smooth scalar defect obey \(\delta\alpha_{xy}^{\rm tr}/\delta\sigma_{xy}=(E_c-\mu)/(eT)\).  At the retained long-wavelength order, the orbital form factor, defect geometry, and common tip kernel cancel after the heat-magnetization current is removed, so the zero of the thermoelectric contrast is pinned by energy weighting at \(E_c=\mu\) rather than by defect shape.  In the hydrodynamic regime, the measurable \(q^2\) tensor amplitudes mix Hall, longitudinal, transverse, boundary, electrothermal, and kinetic channels, so a Hall-odd image is not by itself a Hall-viscosity measurement.  For a representative graphene geometry, a Schur-complement fit against the stated nuisance library yields a conditional one-standard-deviation sensitivity \(\delta d_H\simeq68\,\mathrm{nm}^2\) at \({\rm SNR}_0=200\), with boundary slip the limiting nuisance.  The framework turns visual interpretation of quantum Hall nanoscopy into a quantitative observability test for electrical, thermoelectric, and viscous response channels.
\end{abstract}
\maketitle

\section{Introduction and conceptual overview}
\label{sec:introduction}

A scanning probe is a mechanical reader of fields, not of coefficients.  It injects or perturbs a device, lets charge, heat, and momentum redistribute through the sample, and then records a spatially blurred scalar.  In a magnetic field the blur is not merely instrumental.  The electron first moves on a cyclotron orbit, so a scalar defect is projected by a Landau-level form factor.  A thermal current contains a circulating magnetization part that never carries charge between contacts.  A viscous stress produces a Hall-odd nonlocal pattern that can be mimicked by boundary slip or by a truncated kinetic closure.  The image is therefore a projection of a response operator, not a photograph of a local material constant.

The operator viewpoint is compact.  Let \(S_{\rm src}\) denote an applied source, \(W_P\) the weighting functional of probe channel \(P\), and \(\calL_{\mathcal R}(B)\) the linear transport operator in regime \(\mathcal R\) at magnetic field \(B\).  The measured contrast has the form
\begin{equation}
        M_{P\leftarrow {\rm src}}^{(\mathcal R)}(B)
        =
        \langle W_P|\calL_{\mathcal R}^{-1}(B)|S_{\rm src}\rangle .
        \label{eq:source_probe}
\end{equation}
Here the brackets denote the spatial and tensor contraction appropriate to the probe.  Equation~\eqref{eq:source_probe} is the continuum version of reciprocity, adjoint-field, injectivity, emissivity, and Shockley--Ramo constructions~\cite{Buttiker1986,Gasparian1996,Gramespacher1997,SongLevitov2014}.  Its role in this paper is not formal decoration.  It fixes the boundary between what is measured and what is inferred: the experiment measures \(M\); transport coefficients enter only after a forward model and an inverse problem have been specified.

The graphene geometry used as a concrete test case is a two-dimensional channel in perpendicular field, with magnetic length
\begin{equation}
        \ell_B=\left(\frac{\hbar}{eB}\right)^{1/2},
        \qquad e>0,
        \label{eq:magnetic_length}
\end{equation}
where the electron charge is \(-e\).  A smooth circular defect or gate-defined junction of radius \(a\) is imaged by a probe of height \(h\) and spot size \(s\).  Such length scales are comparable to those in graphene quantum Hall nanoscopy and viscous-flow measurements~\cite{Sulpizio2019,Ku2020,Jenkins2022,Krebs2026}.  The external Hamiltonian potential is denoted \(U(\rr)\).  It is not the same object as the induced electrochemical potential, the temperature perturbation, the current field, or a hydrodynamic velocity.

Two quantitative consequences are developed.  The first is microscopic and topological.  A smooth defect projected to Landau level \(N\) enters as \(F_N(q)U(\qv)\bar\rho_{-\qv}\), not as \(U(\qv)\rho_{-\qv}\).  The same projected defect and the same tip kernel multiply the defect-induced electrical Hall response and the magnetization-subtracted thermoelectric Hall response.  For a sharp mobility edge at energy \(E_c\), the common factors cancel at the retained long-wavelength order, giving
\begin{equation}
        \frac{\delta\alpha_{xy}^{\rm tr}(\qv)}{\delta\sigma_{xy}(\qv)}
        =
        \frac{E_c-\mu}{eT} .
        \label{eq:thermal_ratio_intro}
\end{equation}
The zero of the thermoelectric contrast is therefore pinned by energy weighting at \(E_c=\mu\), not by the defect shape.

The second consequence is mesoscopic and statistical.  A Hall-viscous coefficient \(d_H=\eta_H/\Gamma\)~\cite{Avron1995,HoyosSon2012} contributes to the \(q^2\) part of a finite-wave-vector conductivity tensor, but a Hall-odd image shape alone is not a measurement of \(d_H\).  Compressible stiffness, electrothermal pressure, even-viscous shear, boundary leakage, and higher-gradient kinetic terms can overlap with it over the finite wave-number band transmitted by the tip.  The extraction is therefore a Fisher problem.  For the default graphene geometry
\begin{equation}
        (\ell_B,a,h,s,\zeta)=(10,80,30,20,50)\,\mathrm{nm},
        \qquad
        \beta=\omega_c\tau_{\rm mr}=1,
        \label{eq:default_geometry_intro}
\end{equation}
where \(\zeta\) is a boundary-slip length and \(\tau_{\rm mr}\) a momentum-relaxation time, the Schur-marginalized error is
\begin{equation}
        \delta d_H\simeq68\,\mathrm{nm}^2
        \qquad
        ({\rm SNR}_0=200).
        \label{eq:dh_intro}
\end{equation}
The projected nuisance space removes about \(99\%\) of the raw Hall-odd Fisher weight; the residual component sets the conditional sensitivity in Eq.~\eqref{eq:dh_intro}.  The point \(\beta=1\) is not a numerical accident: it is the tensor-orthogonal point at which Hall-viscous and even longitudinal/transverse directions decouple before any radial fitting.

These two results are complementary, not simultaneous.  The thermoelectric Hall ratio is a statement about the strong-field, sharp-Landau-level regime, where transport is activated and the mobility edge is well defined.  The Hall-viscometry bound is a statement about the hydrodynamic regime at \(\beta=\omega_c\tau_{\rm mr}\sim1\), where momentum-conserving collisions generate the nonlocal \(q^2\) response.  The two are accessed in different windows of \((B,T,\mu)\) on the same platform, and the response-regime bookkeeping of Appendix~\ref{app:response_matrix} exists precisely to keep them from being compared as if they were one measurement.

\begin{figure}[t]
\centering
\resizebox{0.95\columnwidth}{!}{%
\begin{tikzpicture}[font=\small, node distance=7pt]
\tikzset{box/.style={draw, rounded corners=2pt, align=center, inner sep=4pt, minimum height=0.65cm}, arrow/.style={-{Latex[length=2mm]}, thick}}
\node[box] (source) {source\\$S_{\rm src}$};
\node[box, right=of source] (operator) {transport\\$\calL_{\mathcal R}^{-1}(B)$};
\node[box, right=of operator] (probe) {probe\\$W_P$};
\draw[arrow] (source) -- (operator);
\draw[arrow] (operator) -- (probe);
\node[box, below=12pt of operator, text width=0.82\columnwidth] (factors) {forward factors: $F_N(q)U(\qv)$, $\calT_{h,s}(q)$, energy weighting, local-background null, nuisance projection};
\draw[arrow] (operator.south) -- (factors.north);
\node[box, below left=10pt and -5pt of factors, text width=0.38\columnwidth] (thermal) {thermoelectric Hall ratio\\Eq.~\eqref{eq:triangle_ratio}};
\node[box, below right=10pt and -5pt of factors, text width=0.38\columnwidth] (schur) {Hall-viscous Schur fit\\Eq.~\eqref{eq:dH_error}};
\draw[arrow] (factors.south west) -- (thermal.north);
\draw[arrow] (factors.south east) -- (schur.north);
\end{tikzpicture}%
}
\caption{Source--operator--probe structure of the paper.  A scanning experiment measures a finite-resolution functional of a linear transport operator.  The same forward factors cancel at the retained order in the thermoelectric Hall ratio or set the Fisher metric that determines whether a Hall-viscous coefficient is identifiable.}
\label{fig:roadmap}
\end{figure}
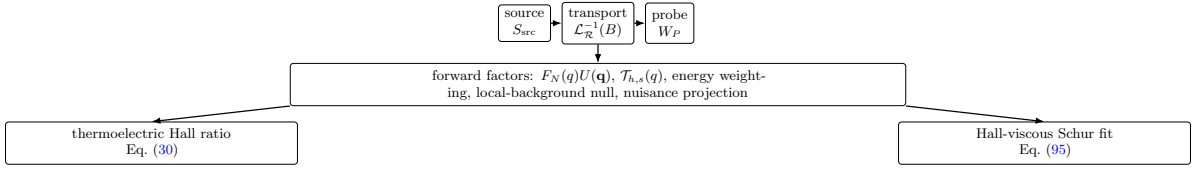

Figure~\ref{fig:roadmap} shows the logic.  Section~\ref{sec:microscopic} builds the microscopic projection and conservation identities.  Section~\ref{sec:hydro} derives the hydrodynamic tensor and kinetic truncation.  Section~\ref{sec:observability} turns finite-resolution images into Fisher information and Schur-projected error bars.  Section~\ref{sec:extensions} states the controlled geometric, finite-frequency, and nonlocal thermoelectric extensions.

\section{Microscopic foundations and conservation identities}
\label{sec:microscopic}

At the microscopic scale the probe does not push on a point electron.  It pushes on a cyclotron orbit.  That orbit has internal structure, and the density operator splits into an orbital form factor and a guiding-centre density.  This is the first correction to the naive picture.  The second correction is subtler: a Kubo bubble built from a heat current counts circulating equilibrium currents as if they were transport.  The subtraction of these magnetization currents is not a convention; it is the step that restores local conservation and removes a bulk-edge ambiguity.

\subsection{Projected defect vertex}
\label{subsec:projected_vertex}

The Fourier convention is
\begin{equation}
        f(\rr)=\intq f(\qv)\ee^{\ii\qv\cdot\rr},
        \qquad
        \intq\equiv\int\frac{\dd^2q}{(2\pi)^2}.
        \label{eq:fourier}
\end{equation}
A weak scalar potential \(U(\rr)\) enters the Hamiltonian as
\begin{equation}
        H_U=\intq U(\qv)\rho_{-\qv},
        \label{eq:HU}
\end{equation}
where \(\rho_{\qv}\) is the full electronic density.  In a magnetic field the position decomposes into a cyclotron coordinate and a guiding-centre coordinate.  Projection to Landau level \(N\) gives
\begin{equation}
        P_N\rho_{\qv}P_N
        =
        F_N(q)\bar\rho_{\qv},
        \label{eq:density_projection}
\end{equation}
where \(P_N\) is the Landau-level projector, \(\bar\rho_{\qv}\) is the guiding-centre density, and \(F_N(q)\) is the orbital form factor.  Hence
\begin{equation}
        P_NH_UP_N
        =
        \intq U(\qv)F_N(q)\bar\rho_{-\qv}.
        \label{eq:projected_vertex}
\end{equation}
For a parabolic-band Landau level,
\begin{equation}
        F_N(q)=L_N(x)\ee^{-x/2},
        \qquad
        x=\frac{q^2\ell_B^2}{2}.
        \label{eq:parabolic_form_factor}
\end{equation}
For monolayer graphene the Dirac spinor mixes adjacent oscillator indices.  For \(N\ge1\),
\begin{equation}
        F_N^{\rm gr}(q)
        =
        \frac{1}{2}\left[L_N(x)+L_{N-1}(x)\right]\ee^{-x/2},
        \qquad N\ge1,
        \label{eq:graphene_form_factor}
\end{equation}
whereas the zero mode has only one oscillator component,
\begin{equation}
        F_0^{\rm gr}(q)=\ee^{-x/2} .
        \label{eq:graphene_zero_mode}
\end{equation}
Equations~\eqref{eq:graphene_form_factor} and \eqref{eq:graphene_zero_mode} are the required separate treatment of higher Dirac Landau levels and the zero mode~\cite{GirvinJach1984,GMP1986,Goerbig2011}.  Expanding at small wave number gives
\begin{align}
        F_N^{\rm gr}(q)&=1-\frac{N}{2}q^2\ell_B^2+\order(q^4\ell_B^4),\qquad N\ge1,\label{eq:smallq_higher}\\
        F_0^{\rm gr}(q)&=1-\frac{1}{4}q^2\ell_B^2+\order(q^4\ell_B^4).\label{eq:smallq_zero}
\end{align}
The assumptions behind this vertex are explicit: the perturbation is smooth, \(q\ell_B\ll1\), and weak enough to avoid Landau-level mixing, \(|U_0|\ll\hbar\omega_c\) in a parabolic estimate.

For the circular defect used in the numerical forward model,
\begin{equation}
        U(\rr)=U_0\Theta(a-r),
        \qquad
        U(q)=2\pi U_0a\frac{J_1(qa)}{q},
        \label{eq:circular_defect}
\end{equation}
where \(a\) is the defect radius.  The projected graphene profile is
\begin{equation}
        U_N^{\rm gr}(R)
        =
        U_0a\int_0^\infty\dd q\,J_1(qa)J_0(qR)F_N^{\rm gr}(q),
        \label{eq:projected_profile}
\end{equation}
with \(R=|\RR|\).  This profile is a known forward input, not an unknown coefficient to be extracted.

\subsection{Ward identity for the defect triangle}
\label{subsec:ward}

The defect correction to a two-current response is a connected triangle with one density insertion:
\begin{equation}
        \delta\Pi_{AB}(\qv)
        =
        -U(\qv)F_N(q)
        \int_0^{1/T}\dd\tau\,
        \langle T_\tau A(\tau)B(0)\bar\rho_{-\qv}(0)\rangle_c .
        \label{eq:triangle}
\end{equation}
Here \(A\) and \(B\) are current operators, \(T_\tau\) orders imaginary time, and the subscript \(c\) denotes the connected part.  At \(\qv\to0\), the operator \(\bar\rho_{-\qv}\) becomes the total particle number.  A uniform scalar insertion is therefore equivalent to shifting the chemical potential.  This statement is the particle-number Ward identity.

To see the sign and derivative explicitly, write the thermal expectation value of an operator product \(X\) as
\begin{equation}
        \langle X\rangle_\mu
        =
        \frac{\mathrm{Tr}\,\ee^{-\beta_T(H-\mu N)}X}{\mathrm{Tr}\,\ee^{-\beta_T(H-\mu N)}},
        \qquad
        \beta_T=\frac{1}{T}.
        \label{eq:thermal_average}
\end{equation}
Differentiating with respect to \(\mu\) gives
\begin{equation}
        \partial_\mu\langle X\rangle_\mu
        =
        \int_0^{\beta_T}\dd\tau\,
        \langle T_\tau N(\tau)X(0)\rangle_c .
        \label{eq:ward_derivative}
\end{equation}
A positive scalar potential raises the local electron energy by \(U\), so the induced spectral shift is \(\delta\eps=U\) and \(-\partial_\mu=\partial_\eps\).  For the Hall response written as
\begin{equation}
        \sigma_{xy}
        =
        \int\dd\eps\,[-f_0'(\eps)]\Phi(\eps),
        \qquad
        f_0(\eps)=\frac{1}{\ee^{(\eps-\mu)/T}+1},
        \label{eq:sigma_spectral}
\end{equation}
where \(\Phi(\eps)\) is the zero-temperature Hall spectral weight, Eq.~\eqref{eq:ward_derivative} implies
\begin{equation}
        \delta\Phi(\eps,\qv)
        =
        -F_N(q)U(\qv)\partial_\eps\Phi(\eps)
        +\order(q^2\ell_B^2).
        \label{eq:spectral_shift}
\end{equation}
The Ward identity fixes the uniform \(q=0\) insertion.  Equation~\eqref{eq:spectral_shift} additionally uses the local spectral-shift approximation for a slowly varying defect; the \(\order(q^2\ell_B^2)\) term collects nonuniform-response corrections beyond that limit.

\subsection{Magnetization-current subtraction and thermoelectric Hall ratio}
\label{subsec:thermal_ratio}

A heat-current bubble in a magnetic field is treacherous because a thermal gradient can excite currents that circulate locally and never cross a contact.  Those persistent magnetization currents are real microscopic currents, but they are not transport.  Counting them as transport gives a spurious edge contribution: the bubble remains nonzero at the mobility edge where the transport thermal contrast must change sign.

For a sharp mobility edge,
\begin{equation}
        \Phi(\eps)=\Phi_0\Theta(\eps-E_c),
        \qquad
        t=\frac{E_c-\mu}{T},
        \qquad
        g(t)=\frac{1}{4\cosh^2(t/2)},
        \label{eq:edge_defs}
\end{equation}
where \(\Phi_0\) is the signed jump in Hall spectral weight.  Combining Eqs.~\eqref{eq:sigma_spectral} and \eqref{eq:spectral_shift} gives
\begin{equation}
        \delta\sigma_{xy}(\qv)
        =
        -F_N(q)U(\qv)\frac{\Phi_0}{T}g(t).
        \label{eq:delta_sigma}
\end{equation}
The magnetization-subtracted transport thermoelectric Hall coefficient is
\begin{equation}
        \alpha_{xy}^{\rm tr}
        =
        \frac{1}{eT}\int\dd\eps\,(\eps-\mu)[-f_0'(\eps)]\Phi(\eps).
        \label{eq:alpha_transport}
\end{equation}
Equivalently, after integrating by parts,
\begin{equation}
        \alpha_{xy}^{\rm tr}
        =
        \frac{1}{e}\int\dd\eps\,s(\eps)\partial_\eps\Phi(\eps),
        \label{eq:entropy_form}
\end{equation}
where \(s(\eps)=-f_0\ln f_0-(1-f_0)\ln(1-f_0)\) is the entropy per state.  Equation~\eqref{eq:alpha_transport} fixes the sign convention for \(\alpha_{xy}^{\rm tr}\); adopting the opposite convention reverses the sign of every quoted \(\alpha\) and of Eq.~\eqref{eq:triangle_ratio}, but not the zero at \(E_c=\mu\).  The entropy form makes explicit that the equilibrium magnetization ambiguity has been removed~\cite{Luttinger1964,JonsonGirvin1984,CooperHalperinRuzin1997,QinNiuShi2011}.

Inserting Eq.~\eqref{eq:spectral_shift} into Eq.~\eqref{eq:alpha_transport} gives
\begin{equation}
        \delta\alpha_{xy}^{\rm tr}(\qv)
        =
        -F_N(q)U(\qv)\frac{\Phi_0}{eT}t g(t).
        \label{eq:delta_alpha_tr}
\end{equation}
The cancellation of the spurious term can be written at the same level.  With the sign convention of Eq.~\eqref{eq:delta_sigma}, the bare heat-current bubble and the magnetization correction are
\begin{align}
        \delta\widetilde\alpha_{xy}^{K}(\qv)
        &=
        -F_N(q)U(\qv)\frac{\Phi_0}{eT}\,[t g(t)-f(t)],
        \label{eq:bare_alpha}\\
        \delta\alpha_{xy}^{M}(\qv)
        &=
        -F_N(q)U(\qv)\frac{\Phi_0}{eT}\,f(t),
        \label{eq:mag_alpha}
\end{align}
where \(f(t)=1/(\ee^t+1)\).  Their sum is
\begin{equation}
        \delta\alpha_{xy}^{\rm tr}
        =
        \delta\widetilde\alpha_{xy}^{K}+\delta\alpha_{xy}^{M}
        =
        -F_N(q)U(\qv)\frac{\Phi_0}{eT}t g(t).
        \label{eq:alpha_sum}
\end{equation}
The \(f(t)\) term is precisely the unphysical zeroth-moment contribution.  It is nonzero at \(t=0\) in the bare bubble and cancels only after the magnetization current is included.  Dividing Eq.~\eqref{eq:delta_alpha_tr} by Eq.~\eqref{eq:delta_sigma} gives the probe-independent ratio
\begin{equation}
        \boxed{
        \frac{\delta\alpha_{xy}^{\rm tr}(\qv)}{\delta\sigma_{xy}(\qv)}
        =
        \frac{t}{e}
        =
        \frac{E_c-\mu}{eT}.}
        \label{eq:triangle_ratio}
\end{equation}
At the retained long-wavelength order, all three common geometric factors--the orbital form factor, the defect profile, and the tip transfer function--cancel.  For a broadened mobility edge, the common factor still separates under the same local spectral-shift approximation:
\begin{align}
        \delta\sigma_{xy}(\qv)&=F_N(q)U(\qv)\chi_V,\label{eq:chiV}\\
        \delta\alpha_{xy}^{\rm tr}(\qv)&=F_N(q)U(\qv)\chi_{\rm th},\label{eq:chith}
\end{align}
with
\begin{align}
        \chi_V&=-\int\dd\eps\,[-f_0'(\eps)]\partial_\eps\Phi(\eps),\label{eq:chiV_def}\\
        \chi_{\rm th}&=-\frac{1}{eT}\int\dd\eps\,[-f_0'(\eps)](\eps-\mu)\partial_\eps\Phi(\eps).\label{eq:chith_def}
\end{align}
The subscript \({\rm th}\) labels a thermal or photothermal response channel; it is not the temperature \(T\).

\begin{figure}[t]
\centering
\includegraphics[width=0.85\columnwidth]{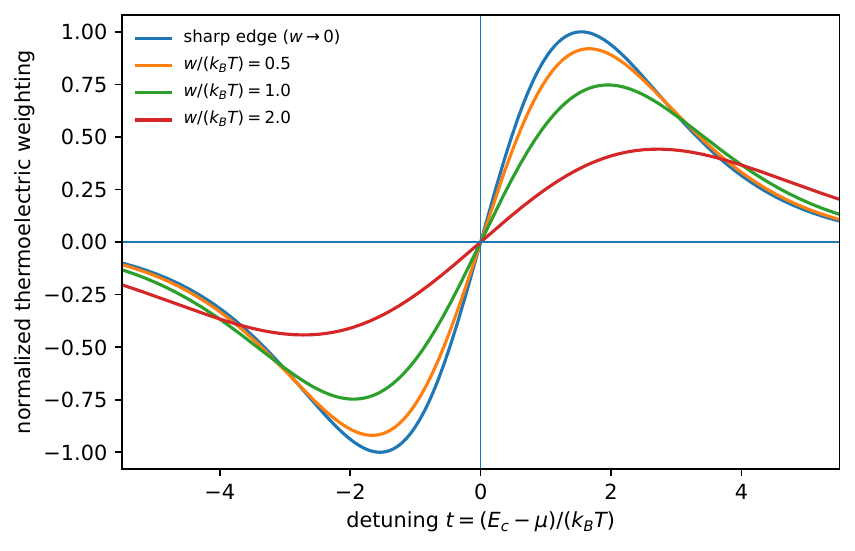}
\caption{Thermoelectric Hall contrast for a mobility edge of finite spectral width.  The horizontal axis is the dimensionless detuning \(t=(E_c-\mu)/T\).  The vertical axis is the thermoelectric defect response normalized to the sharp-edge peak.  The sharp edge gives \(t g(t)\), with a zero fixed at \(E_c=\mu\).  Symmetric broadening lowers the peak and shifts the extrema outward while preserving the sign change.}
\label{fig:thermal_filter}
\end{figure}

\section{Mesoscopic continuum and non-local hydrodynamic operators}
\label{sec:hydro}

At mesoscopic scales an electron fluid does not simply stop where the local conductivity is low.  Momentum carried by neighbouring fluid elements pushes shear stress into the depleted region and heals the flow over a viscous length.  This is the mechanical origin of viscous or kinetic erasure: a local Ohmic map would show a sharply suppressed current region, while a nonlocal stress term spreads momentum and reduces the contrast~\cite{Torre2015,LevitovFalkovich2016,Alekseev2016}.  In Fourier space that healing is the appearance of powers of \(q^2\) in the inverse transport operator.

\subsection{Wave-vector-adapted Stokes--Ohm matrix}
\label{subsec:hydro_matrix}

Let
\begin{equation}
        \hat{\mathbf e}_L=\hat\qv=\frac{\qv}{q},
        \qquad
        \hat{\mathbf e}_\perp=\hat{\mathbf z}\times\hat\qv .
        \label{eq:basis}
\end{equation}
For a velocity field \(\mathbf u\), define \(u_L=\hat{\mathbf e}_L\cdot\mathbf u\) and \(u_\perp=\hat{\mathbf e}_\perp\cdot\mathbf u\), with analogous definitions for \(E_L\) and \(E_\perp\).  In this basis an isotropic magneto-viscous response has only one antisymmetric entry.  The linearized Stokes--Ohm or magneto-Brinkman balance is
\begin{equation}
        A(q)
        \begin{pmatrix}
        u_L\\ u_\perp
        \end{pmatrix}
        =
        ne
        \begin{pmatrix}
        E_L\\ E_\perp
        \end{pmatrix},
        \qquad
        A(q)=
        \begin{pmatrix}
        A_L & C\\
        -C & A_\perp
        \end{pmatrix}.
        \label{eq:A_matrix}
\end{equation}
Here \(n\) is the carrier density.  The momentum susceptibility has been absorbed into the coefficients of \(A\).  Thus \(\Gamma\), \(\Omega\), and \(\eta q^2\) have the same units; in a rate-normalized convention \(\Gamma=1/\tau_{\rm mr}\) and \(\Omega=\omega_c\).  To order \(q^2\),
\begin{align}
        A_\perp&=\Gamma+\eta q^2,\label{eq:Aperp}\\
        A_L&=\Gamma+(\eta_L+\Lambda_c)q^2,\label{eq:AL}\\
        C&=\Omega+\eta_H q^2.\label{eq:Centry}
\end{align}
The coefficient \(\eta\) is the transverse shear coefficient, \(\eta_L\) the longitudinal viscous coefficient, \(\eta_H\) the Hall-viscous coefficient~\cite{Pellegrino2017}, and \(\Lambda_c\) a non-viscous longitudinal stiffness collecting compressibility, pressure relaxation, or electrothermal pressure.  Introduce
\begin{equation}
        \beta=\frac{\Omega}{\Gamma}=\omega_c\tau_{\rm mr},
        \qquad
        d_\perp=\frac{\eta}{\Gamma},
        \qquad
        d_L=\frac{\eta_L+\Lambda_c}{\Gamma},
        \qquad
        d_H=\frac{\eta_H}{\Gamma}.
        \label{eq:d_definitions}
\end{equation}
All three \(d\)'s have dimensions of length squared.  In the simplest shear-only limit, \(D_v=d_\perp^{1/2}=(\eta/\Gamma)^{1/2}\) is the Gurzhi or viscous healing length.

\subsection{Gradient expansion and tensor amplitudes}
\label{subsec:gradient_expansion}

The current is \(\mathbf j=ne\mathbf u\), so
\begin{equation}
        \sigma(q)=(ne)^2A^{-1}(q).
        \label{eq:sigma_Ainv}
\end{equation}
The inverse of the matrix in Eq.~\eqref{eq:A_matrix} is
\begin{equation}
        A^{-1}
        =
        \frac{1}{\det A}
        \begin{pmatrix}
        A_\perp & -C\\
        C & A_L
        \end{pmatrix},
        \qquad
        \det A=A_LA_\perp+C^2 .
        \label{eq:A_inverse}
\end{equation}
Therefore
\begin{align}
        \sigma_{LL}(q)&=\frac{(ne)^2A_\perp}{\det A},\label{eq:sigmaLL_exact}\\
        \sigma_{\perp\perp}(q)&=\frac{(ne)^2A_L}{\det A},\label{eq:sigmaperp_exact}\\
        \sigma_H(q)&\equiv-\sigma_{\perp L}(q)=-\frac{(ne)^2C}{\det A}.\label{eq:sigmaH_exact}
\end{align}
Set
\begin{equation}
        a_L=\eta_L+\Lambda_c,
        \qquad
        a_\perp=\eta,
        \qquad
        c=\eta_H .
        \label{eq:abc_defs}
\end{equation}
Substituting Eqs.~\eqref{eq:Aperp}--\eqref{eq:Centry} into the determinant gives
\begin{align}
        \det A
        &=
        (\Gamma+a_Lq^2)(\Gamma+a_\perp q^2)+(\Omega+cq^2)^2
        \nonumber\\
        &=
        \Gamma^2+\Omega^2
        +q^2[\Gamma(a_L+a_\perp)+2\Omega c]
        +\order(q^4)
        \nonumber\\
        &\equiv
        D_0+q^2D_2+\order(q^4),
        \label{eq:det_expansion}
\end{align}
where
\begin{equation}
        D_0=\Gamma^2(1+\beta^2),
        \qquad
        D_2=\Gamma(a_L+a_\perp)+2\Omega c .
        \label{eq:D0D2}
\end{equation}
The reciprocal determinant is
\begin{equation}
        \frac{1}{\det A}
        =
        \frac{1}{D_0}
        \left(1-q^2\frac{D_2}{D_0}\right)+\order(q^4).
        \label{eq:inverse_det_expansion}
\end{equation}
Using \(\sigma_D=(ne)^2/\Gamma\), the zeroth-order conductivities are
\begin{equation}
        \sigma_{LL}^{(0)}=\frac{\sigma_D}{1+\beta^2},
        \qquad
        \sigma_H^{(0)}=-\frac{\sigma_D\beta}{1+\beta^2}.
        \label{eq:drude_field}
\end{equation}
The \(q^2\) part of \(\sigma_{LL}\) follows from
\begin{align}
        \sigma_{LL}(q)
        &=
        (ne)^2(\Gamma+a_\perp q^2)
        \frac{1}{D_0}
        \left(1-q^2\frac{D_2}{D_0}\right)+\order(q^4)
        \nonumber\\
        &=
        \frac{(ne)^2\Gamma}{D_0}
        +q^2\frac{(ne)^2}{D_0}
        \left(a_\perp-\Gamma\frac{D_2}{D_0}\right)
        +\order(q^4).
        \label{eq:sigmaLL_expand_steps}
\end{align}
Since
\begin{equation}
        \Gamma\frac{D_2}{D_0}
        =
        \frac{a_L+a_\perp+2\beta c}{1+\beta^2},
        \label{eq:GammaD2D0}
\end{equation}
this gives
\begin{equation}
        \sigma_{LL}^{(2)}
        =
        \frac{\sigma_D}{(1+\beta^2)^2}
        \left(\beta^2d_\perp-d_L-2\beta d_H\right).
        \label{eq:sigmaLL2}
\end{equation}
The transverse component is expanded in the same determinant without using a separate assumption:
\begin{align}
        \sigma_{\perp\perp}(q)
        &=
        (ne)^2(\Gamma+a_L q^2)
        \frac{1}{D_0}
        \left(1-q^2\frac{D_2}{D_0}\right)+\order(q^4)
        \nonumber\\
        &=
        \frac{(ne)^2\Gamma}{D_0}
        +q^2\frac{(ne)^2}{D_0}
        \left(a_L-\Gamma\frac{D_2}{D_0}\right)
        +\order(q^4),
        \label{eq:sigmaperp_expand_steps}
\end{align}
so that
\begin{equation}
        \sigma_{\perp\perp}^{(2)}
        =
        \frac{\sigma_D}{(1+\beta^2)^2}
        \left(\beta^2d_L-d_\perp-2\beta d_H\right).
        \label{eq:sigmaperp2}
\end{equation}
The Hall component is expanded as
\begin{align}
        \sigma_H(q)
        &=
        -(ne)^2(\Omega+cq^2)
        \frac{1}{D_0}
        \left(1-q^2\frac{D_2}{D_0}\right)+\order(q^4)
        \nonumber\\
        &=
        -\frac{(ne)^2\Omega}{D_0}
        -q^2\frac{(ne)^2}{D_0}
        \left(c-\Omega\frac{D_2}{D_0}\right)+\order(q^4).
        \label{eq:sigmaH_expand_steps}
\end{align}
With
\begin{equation}
        \Omega\frac{D_2}{D_0}
        =
        \frac{\beta(a_L+a_\perp+2\beta c)}{1+\beta^2},
        \label{eq:OmegaD2D0}
\end{equation}
the normalized length-squared amplitudes
\begin{equation}
        S_{ij}=\frac{(1+\beta^2)^2}{\sigma_D}\sigma_{ij}^{(2)},
        \qquad ij\in\{LL,\perp\perp,H\},
        \label{eq:S_definition}
\end{equation}
are
\begin{align}
        S_{LL}&=\beta^2d_\perp-d_L-2\beta d_H,\label{eq:SLL}\\
        S_{\perp\perp}&=\beta^2d_L-d_\perp-2\beta d_H,\label{eq:Sperp}\\
        S_H&=\beta(d_L+d_\perp)-(1-\beta^2)d_H.\label{eq:SH}
\end{align}
Solving this linear system gives
\begin{align}
        d_H&=\frac{(\beta^2-1)S_H-\beta(S_{LL}+S_{\perp\perp})}{(1+\beta^2)^2},\label{eq:dH_inverse}\\
        d_L&=\frac{2\beta S_H-S_{LL}+\beta^2S_{\perp\perp}}{(1+\beta^2)^2},\label{eq:dL_inverse}\\
        d_\perp&=\frac{2\beta S_H+\beta^2S_{LL}-S_{\perp\perp}}{(1+\beta^2)^2}.\label{eq:dperp_inverse}
\end{align}
Equation~\eqref{eq:dH_inverse} is the algebraic warning at the centre of Hall viscometry: the Hall-viscous length \(d_H\) is not equal to one Hall-like image coefficient.  It is a particular projection of the full triple \((S_{LL},S_{\perp\perp},S_H)\).

\subsection{Tensor directions and structural orthogonality}
\label{subsec:orthogonality}

The columns of Eqs.~\eqref{eq:SLL}--\eqref{eq:SH} define three tensor directions in the measured basis:
\begin{align}
        \mathbf v_H&=(-2\beta,-2\beta,\beta^2-1),\label{eq:vH}\\
        \mathbf v_L&=(-1,\beta^2,\beta),\label{eq:vL}\\
        \mathbf v_\perp&=(\beta^2,-1,\beta).\label{eq:vperp}
\end{align}
Their overlaps are
\begin{align}
        \mathbf v_H\cdot\mathbf v_L
        &=
        (-2\beta)(-1)+(-2\beta)\beta^2+(\beta^2-1)\beta
        \nonumber\\
        &=
        2\beta-2\beta^3+\beta^3-\beta
        =
        \beta(1-\beta^2),
        \label{eq:vHvL}\\
        \mathbf v_H\cdot\mathbf v_\perp
        &=
        (-2\beta)\beta^2+(-2\beta)(-1)+(\beta^2-1)\beta
        \nonumber\\
        &=
        -2\beta^3+2\beta+\beta^3-\beta
        =
        \beta(1-\beta^2).
        \label{eq:vHvperp}
\end{align}
Thus the strict tensor orthogonality condition is
\begin{equation}
        \mathbf v_H\cdot\mathbf v_L=
        \mathbf v_H\cdot\mathbf v_\perp=0
        \qquad\Longleftrightarrow\qquad
        \beta=0\ \hbox{or}\ |\beta|=1 .
        \label{eq:orthogonality_condition}
\end{equation}
The nontrivial operating point is \(|\beta|=1\).  At \(\beta=1\),
\begin{equation}
        \mathbf v_H=(-2,-2,0),
        \quad
        \mathbf v_L=(-1,1,1),
        \quad
        \mathbf v_\perp=(1,-1,1),
        \label{eq:beta1_vectors}
\end{equation}
and the even longitudinal/transverse channels decouple from the Hall-viscous target by tensor geometry alone.  Away from \(\beta=1\), the even-channel overlap is finite and must be removed statistically.  This is precisely what the Schur projection in Section~\ref{sec:observability} does.

\subsection{Kinetic harmonic closure}
\label{subsec:kinetic}

The hydrodynamic gradient expansion is the low-multipole limit of a kinetic distribution.  As temperature decreases or magnetic field increases, the non-equilibrium distribution on the Fermi surface grows higher angular harmonics; a local viscous closure fails when those harmonics stop being slaved to density, momentum, and stress.

Write the angular distribution as \(f(\theta)=\sum_m f_m\ee^{\ii m\theta}\).  The linearized relaxation-time hierarchy is
\begin{equation}
        \frac{v_F}{2}
        \left[(\partial_x-\ii\partial_y)f_{m-1}
        +(\partial_x+\ii\partial_y)f_{m+1}\right]
        +(\ii m\omega_c+\gamma_m)f_m
        =S_m .
        \label{eq:boltzmann_hierarchy}
\end{equation}
Here \(v_F\) is the Fermi velocity, \(\omega_c\) the signed cyclotron frequency, \(\gamma_m\) the relaxation rate of harmonic \(m\), and \(S_m\) a source.  The sectors \(m=0\) and \(m=\pm1\) carry density and momentum.  The sectors \(m=\pm2\) carry shear and Hall viscosity.  Eliminating the stress harmonic gives the leading nonlocal kinetic length
\begin{equation}
        \ell_K^2(B)
        =
        \frac{v_F^2}{4(\gamma_1+\ii\omega_c)(\gamma_2+2\ii\omega_c)}
        +\order(q^2).
        \label{eq:kinetic_length}
\end{equation}
The \(q^2\) coefficient is therefore the stress-sector response; harmonics \(|m|\ge3\) first modify the current at order \(q^4\), with a generally complex coefficient at finite frequency.  In numerical use one truncates at \(|m|\le M\) and increases \(M\) until both the forward fields and their probe-projected maps obey
\begin{equation}
        \frac{\|M_{M+1}-M_M\|_2}{\|M_M\|_2}\le10^{-3}.
        \label{eq:M_convergence}
\end{equation}
To avoid an accidental single-cutoff crossing, Eq.~\eqref{eq:M_convergence} is required for two successive increments.  The smallest \(M\) satisfying this criterion defines a closure-failure diagnostic \(M(T,B)\).  Growth of \(M(T,B)\) is the operational boundary beyond which a local hydrodynamic description is no longer sufficient.

\section{Observability, inversion, and information metrics}
\label{sec:observability}

The microscope imposes a bandwidth.  A tip at height \(h\) and spot size \(s\) cannot transmit arbitrary wave number.  Trying to divide the image by that transfer function would amplify the empty tail of the measurement.  The correct procedure is the opposite: keep the transfer function in the forward model and ask which parameter directions remain distinguishable inside the transmitted band.

\subsection{Finite-tip transfer and scalar Fisher warm-up}
\label{subsec:fisher_warmup}

For an electrostatic tip, the dimensionless transfer function is
\begin{equation}
        \calT_{h,s}(q)=\ee^{-qh-q^2s^2/2}.
        \label{eq:tip_transfer}
\end{equation}
The calibrated filtered defect profile is
\begin{equation}
        A_N(q)=\calT_{h,s}(q)F_N^{\rm gr}(q)U(q).
        \label{eq:AN}
\end{equation}
For the default geometry \(\ell_B=10\,\mathrm{nm}\), \(a=80\,\mathrm{nm}\), \(N=1\), \(h=30\,\mathrm{nm}\), and \(s=20\,\mathrm{nm}\), the \(N=1\) graphene form factor vanishes at \(q=2/\ell_B=0.2\,\mathrm{nm}^{-1}\).  The factor \(\ee^{-qh}\) is then \(\ee^{-6}\), so the Fisher integrals are dominated by \(q\lesssim1/h\), well below the form-factor node.

To fix normalization, first consider a scalar image with local background removed:
\begin{equation}
        V(\qv)=A_N(q)\left[\lambda_2q^2+\lambda_4a^2q^4\right]+\xi(\qv).
        \label{eq:scalar_forward}
\end{equation}
The coefficient \(\lambda_2\) is the reported nonlocal length-squared amplitude.  The coefficient \(\lambda_4\) is a higher-gradient nuisance amplitude.  The noise is white in the measured band,
\begin{equation}
        \langle\xi(\qv)\xi^*(\qv')\rangle
        =(2\pi)^2S_V\delta(\qv-\qv'),
        \label{eq:white_noise}
\end{equation}
where \(S_V\) is the image-noise spectral density.  Define radial moments
\begin{equation}
        I_m=\intq |A_N(q)|^2q^m.
        \label{eq:radial_moments}
\end{equation}
For \(\theta=(\lambda_2,\lambda_4)\), the Fisher matrix is
\begin{equation}
        F_{ij}
        =
        \frac{1}{2S_V}
        \intq
        \frac{\partial V^*(\qv)}{\partial\theta_i}
        \frac{\partial V(\qv)}{\partial\theta_j}.
        \label{eq:fisher_scalar_def}
\end{equation}
The factor \(1/2\) accounts for the reality constraint between \(\qv\) and \(-\qv\).  Substitution gives
\begin{equation}
        F=
        \frac{1}{2S_V}
        \begin{pmatrix}
        I_4 & a^2I_6\\
        a^2I_6 & a^4I_8
        \end{pmatrix}.
        \label{eq:fisher_scalar}
\end{equation}
The marginalized variance of \(\lambda_2\) is the \((1,1)\) element of \(F^{-1}\):
\begin{equation}
        {\rm Var}(\hat\lambda_2)
        =
        2S_V\frac{I_8}{I_4I_8-I_6^2}.
        \label{eq:lambda_variance}
\end{equation}
The denominator is the Gram determinant of the two transmitted shapes \(q^2A_N\) and \(a^2q^4A_N\).  When the finite band makes those shapes parallel, the error diverges.  A reference signal-to-noise ratio is
\begin{equation}
        {\rm SNR}_0^2=\frac{I_0}{S_V},
        \qquad
        I_0=\intq |A_N(q)|^2.
        \label{eq:SNR0}
\end{equation}
For the default geometry the scalar inverse gives \(\delta\lambda_2=5.410\times10^3\,\mathrm{nm}^2/{\rm SNR}_0\), or \(27.1\,\mathrm{nm}^2\) at \({\rm SNR}_0=200\).

\subsection{Schur complement and nuisance projection}
\label{subsec:schur}

The Hall-viscous inverse problem uses the measured basis \((S_{LL},S_{\perp\perp},S_H)\).  The radial inner product is
\begin{equation}
        (X|Y)_0
        =
        \int_0^\infty\dd q\,W(q;h,s)X(q)\cdot Y(q),
        \label{eq:inner_product}
\end{equation}
where the dot product is in the three-component tensor basis.  For the circular defect,
\begin{equation}
        W(q;h,s)
        =
        \frac{q}{2\pi}
        \left|
        \calT_{h,s}(q)F_1^{\rm gr}(q)2\pi a\frac{J_1(qa)}{q}
        \right|^2 .
        \label{eq:radial_weight}
\end{equation}
The Hall-viscous target template and the nuisance library are
\begin{align}
        T_H&=q^2\mathbf v_H,\label{eq:TH}\\
        T_c&=q^2\mathbf v_L,\label{eq:Tc}\\
        T_{\rm th}&=q^2[1+(q\ell_{\rm th})^2]^{-1}\mathbf v_L,\label{eq:Tth}\\
        T_\eta&=q^2\mathbf v_\perp,\label{eq:Teta}\\
        T_\zeta&=q^2(1+q\zeta)^{-1}\mathbf v_H,\label{eq:Tzeta}\\
        T_4&=a^2q^4\mathbf v_H.\label{eq:T4}
\end{align}
Here \(T_c\) is compressible stiffness, \(T_{\rm th}\) electrothermal pressure with relaxation length \(\ell_{\rm th}\), \(T_\eta\) even-viscous shear, \(T_\zeta\) Hall-odd boundary-slip leakage with slip length \(\zeta\), and \(T_4\) higher-gradient kinetic closure.  The non-target library is
\begin{equation}
        \calN=\{T_c,T_{\rm th},T_\eta,T_\zeta,T_4\}.
        \label{eq:nuisance_library}
\end{equation}
The slip template is a conservative boundary-leakage basis function.  A Navier boundary condition \(u_\parallel=\zeta\partial_nu_\parallel\) contains the length \(\zeta\); its Fourier-space Robin structure suppresses boundary response once \(q\zeta\gtrsim1\).  The factor \((1+q\zeta)^{-1}\) is therefore a minimal representation of that crossover~\cite{KiselevSchmalian2019}.  It is not a claim that boundary slip is intrinsically Hall odd; only the Hall-odd projection is kept because that is the dangerous component for \(d_H\).

For a single target the Schur complement is
\begin{equation}
        R_H
        =
        (T_H|T_H)_0
        -(T_H|\calN)_0(\calN|\calN)_0^+(\calN|T_H)_0,
        \label{eq:schur}
\end{equation}
where \(+\) is the Moore--Penrose inverse.  In block-matrix notation this is the scalar version of
\begin{equation}
        F^\perp=F_{ss}-F_{sn}F_{nn}^+F_{ns}.
        \label{eq:schur_block}
\end{equation}
The marginalized error is
\begin{equation}
        \delta d_H
        =
        \frac{1}{{\rm SNR}_0}
        \left(\frac{2I_0}{R_H}\right)^{1/2},
        \qquad
        I_0=\int_0^\infty\dd q\,W(q;h,s).
        \label{eq:dH_error}
\end{equation}
For \(\zeta=50\,\mathrm{nm}\), \(\ell_{\rm th}=a\), and \(\beta=1\),
\begin{equation}
        \boxed{
        \delta d_H
        =
        \frac{1.36\times10^4\,\mathrm{nm}^2}{{\rm SNR}_0}
        \simeq68\,\mathrm{nm}^2
        \quad({\rm SNR}_0=200),}
        \label{eq:dH_number}
\end{equation}
with
\begin{equation}
        \frac{R_H}{(T_H|T_H)_0}\simeq1.0\times10^{-2}.
        \label{eq:schur_fraction}
\end{equation}
The projection removes about \(99\%\) of the raw Hall-odd Fisher weight.  The surviving \(1\%\) is the irreducible component on which the error bar is based.

The leave-one-out rule identifies the limiting nuisance.  At \(\beta=1\), \(T_c\), \(T_{\rm th}\), and \(T_\eta\) are tensor-orthogonal to \(T_H\), so removing them leaves \(\delta d_H\) essentially unchanged.  Removing \(T_\zeta\) raises the residual fraction to approximately \(0.5\) and lowers the error to about \(10\,\mathrm{nm}^2\) at \({\rm SNR}_0=200\).  Removing \(T_4\) leaves a smaller but visible degradation.  Boundary slip is therefore the dominant mathematical nuisance controlling bulk Hall-viscosity identifiability.

Sweeping \(\beta\) separates two ideas that are easily confused.  The normalized residual fraction measures degeneracy after the raw target norm is divided out.  The absolute error also depends on
\begin{equation}
        |\mathbf v_H|^2
        =
        (-2\beta)^2+(-2\beta)^2+(\beta^2-1)^2
        =
        \beta^4+6\beta^2+1.
        \label{eq:vH_norm}
\end{equation}
Thus \(\beta=1\) is the clean tensor-decoupling point, not necessarily the global minimum of \(\delta d_H\).  In the reduced model the error decreases at larger \(\beta\) because the raw Hall-viscous tensor norm grows, until the hydrodynamic validity condition \(|\omega_c\tau_{\rm ee}|\ll1\), heating, and device-specific noise set the practical limit.

\begin{figure*}[t]
\centering
\includegraphics[width=0.99\textwidth]{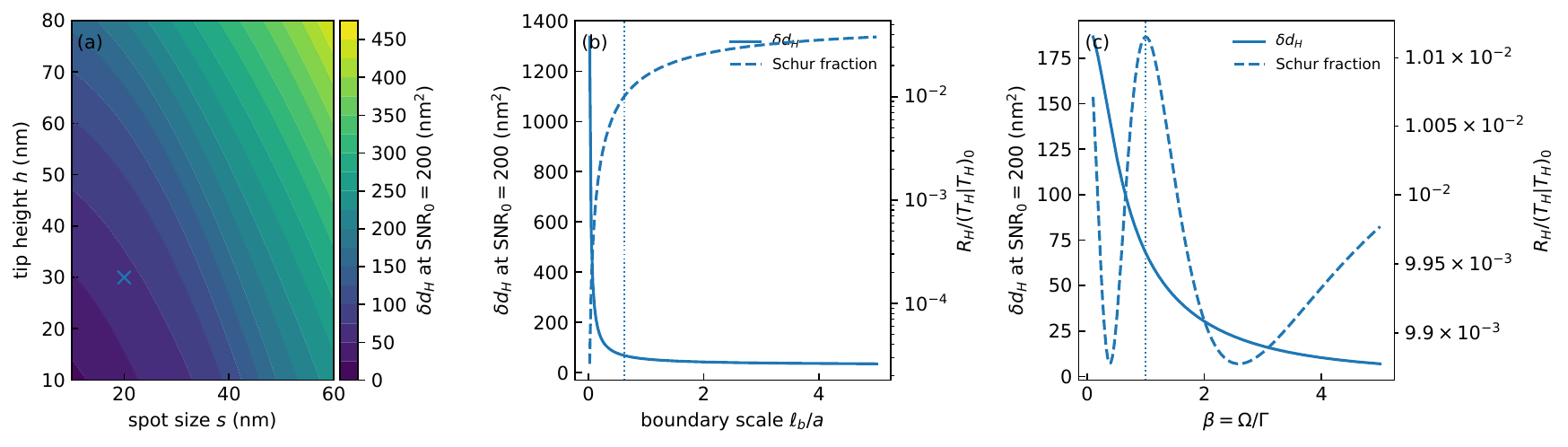}
\caption{Parametric Schur-complement study for Hall-viscous identifiability.  Left: marginalized error \(\delta d_H\) at \({\rm SNR}_0=200\) versus tip height \(h\) and spot size \(s\); larger spatial averaging narrows the transmitted wave-number band and makes radial templates less distinguishable.  Centre: slip-length scan at fixed \(h=30\,\mathrm{nm}\) and \(s=20\,\mathrm{nm}\); the dashed line marks \(\zeta/a=50/80\), and the divergence as \(\zeta\to0\) shows bulk--slip degeneracy.  Right: \(\beta\)-sweep at the default geometry; the Schur fraction stays near \(10^{-2}\), while the absolute error changes because \(|\mathbf v_H|^2=\beta^4+6\beta^2+1\).}
\label{fig:schur_sweeps}
\end{figure*}

\section{Geometric and non-local extensions}
\label{sec:extensions}

The circular defect is the cleanest calibration object.  It is not the only geometry.  The source--operator--probe framework is valuable precisely because the geometry enters through templates and kernels, not through a new interpretation each time.  Three extensions show how the same bookkeeping can be generalized.  They are forward-model formulations, not quantitative predictions for a specific device.

\subsection{Smooth edges and gate-defined junctions}
\label{subsec:edge_extension}

For a soft gate-defined edge, replace the circular profile by a one-dimensional mass or potential wall,
\begin{equation}
        U(x)=\frac{U_0}{2}\left[1+\tanh\left(\frac{x}{w}\right)\right],
        \label{eq:tanh_edge}
\end{equation}
where \(w\) is the edge width.  The derivative localized at the edge is
\begin{equation}
        \partial_xU(x)=\frac{U_0}{2w}\,\mathrm{sech}^2\left(\frac{x}{w}\right).
        \label{eq:edge_derivative}
\end{equation}
For a long straight edge the relevant wave number is \(q_y\) along the boundary and the radial weight of Eq.~\eqref{eq:radial_weight} is replaced by a one-dimensional band weight
\begin{equation}
        W_{\rm edge}(q_y;h,s,w)
        =
        \frac{1}{2\pi}
        \left|
        \calT_{h,s}(q_y)F_N^{\rm gr}(q_y)U_w(q_y)
        \right|^2,
        \label{eq:edge_weight}
\end{equation}
where \(U_w(q_y)\) denotes the Fourier amplitude of the smooth edge mode or of its normal derivative, depending on the probe channel.  A conservative edge-current nuisance template has the form
\begin{equation}
        T_{\rm edge}(q_y)
        =
        q_y^2\,\mathcal E(q_yw,q_y\ell_B)\,\mathbf v_H,
        \label{eq:Tedge}
\end{equation}
with \(\mathcal E\) a dimensionless envelope fixed by the gate profile and edge-state width.  The Schur projection then tests whether bulk \(q^2\mathbf v_H\) is distinguishable from edge-localized Hall-odd transport.

\subsection{Finite-frequency response}
\label{subsec:finite_frequency}

At dc, the dangerous bulk and slip templates can be collinear in tensor space and separated only by radial shape.  A finite drive frequency supplies an additional axis.  In the kinetic hierarchy the replacement is
\begin{equation}
        \gamma_m+\ii m\omega_c
        \longrightarrow
        \gamma_m+\ii m\omega_c-\ii\omega,
        \label{eq:frequency_shift}
\end{equation}
where \(\omega\) is the external drive frequency.  The kinetic length becomes
\begin{equation}
        \ell_K^2(B,\omega)
        =
        \frac{v_F^2}{4(\gamma_1+\ii\omega_c-\ii\omega)(\gamma_2+2\ii\omega_c-\ii\omega)}.
        \label{eq:ellK_frequency}
\end{equation}
Consequently the bulk Hall-viscous coefficient acquires a complex frequency dependence.  In a model where geometric slip leakage is weakly dispersive over the same band, the two channels acquire distinct frequency structure.  The template library becomes
\begin{equation}
        T_H(q,\omega)=q^2\mathbf v_H(\omega),
        \qquad
        T_\zeta(q,\omega)=q^2(1+q\zeta)^{-1}\mathbf v_H(0)+\order(\omega\tau_{\rm edge}).
        \label{eq:frequency_templates}
\end{equation}
The Schur residual \(R_H(\omega)\) is therefore a two-dimensional separation problem in \((q,\omega)\); finite-frequency data can in principle improve identifiability when dc radial shapes are nearly degenerate, but the gain is device and noise-model dependent.

\subsection{Nonlocal thermoelectric response}
\label{subsec:thermoelectric_extension}

Thermal maps probe the energy derivative of transport spectral weight.  In a nonlocal fluid the thermoelectric tensor also has a gradient expansion,
\begin{equation}
        \alpha_{ij}(q)=\alpha_{ij}^{(0)}+q^2\alpha_{ij}^{(2)}+\order(q^4).
        \label{eq:alpha_nonlocal}
\end{equation}
The hydrodynamic Seebeck response can be represented by templates
\begin{equation}
        T_S(q)=q^2\mathbf v_S,
        \qquad
        \mathbf v_S=c_L\mathbf v_L+c_\perp\mathbf v_\perp+c_H\mathbf v_H,
        \label{eq:TS_template}
\end{equation}
where \(c_L\), \(c_\perp\), and \(c_H\) are thermoelectric projection coefficients fixed by the local equation of state and by magnetization-subtracted energy transport.  The thermometry kernel includes energy relaxation,
\begin{equation}
        K_T(q)
        =
        \frac{\ee^{-qh-q^2s^2/2}}{1+q^2\ell_{\rm th}^2},
        \label{eq:thermal_kernel}
\end{equation}
where \(\ell_{\rm th}\) is the thermal relaxation length.  The same Fisher and Schur algebra used for \(d_H\) then provides the metric for asking whether a nonlocal Seebeck or thermoelectric Hall-odd component is observable.

\section{Discussion and conclusion}
\label{sec:conclusion}

The essential distinction is between a contrast and a coefficient.  The factor \(F_N(q)\) is orbital quantum mechanics, \(U(\qv)\) is the calibrated defect, \(\calT_{h,s}(q)\) is the instrument, \(\chi_V\) and \(\chi_{\rm th}\) are energy weightings, \(d_L\), \(d_\perp\), and \(d_H\) are transport lengths, and \(R_H\) is a Schur-projected information weight.  Mixing these objects is how one turns a measured image into an overinterpreted picture.  Keeping them separate turns the same image into a quantitative test.

The microscopic result is the cleanest.  A projected scalar defect shifts the Hall spectral weight by an energy derivative, and magnetization-current subtraction removes the bare-bubble edge artifact.  Within the projected long-wavelength model, the ratio in Eq.~\eqref{eq:triangle_ratio} is independent of the defect shape and the common probe kernel.  It is an experimentally direct check: sweep \(\mu\) or \(T\), compare electrical and thermoelectric Hall contrasts for the same defect, and look for the sign reversal at \(E_c=\mu\).

The viscous result is deliberately more conservative.  It does not assert that a Hall-odd image is Hall viscosity.  It asks whether a Hall-viscous target survives projection against the nuisance library \(\calN\).  For the representative geometry and stated library, the raw information is mostly absorbed by boundary leakage, while the surviving Schur component gives the conditional sensitivity \(\delta d_H\simeq68\,\mathrm{nm}^2\) at \({\rm SNR}_0=200\).  The leave-one-out test identifies boundary slip as the limiting nuisance.

A practical experimental protocol follows directly.  First, calibrate \(U(\qv)\) and apply the Landau-level form factor.  Second, measure or model the actual transfer kernel.  Third, subtract the local \(q^0\) background.  Fourth, quote \({\rm SNR}_0\) in the transmitted band.  Fifth, build a device-specific nuisance library.  Sixth, report the Schur-complement error and the leave-one-out stability.  Only after these steps should a scanning-probe contrast be assigned to a thermal, electrical, or viscous coefficient.

\begin{acknowledgments}
The author thanks M.~M.~Fogler for prior collaboration on inhomogeneous magnetotransport~\cite{ParasharFogler2026PRB} and for the graphene nanoscopy experiments of Z.~J.~Krebs, V.~W.~Brar, and collaborators~\cite{Krebs2026}, which motivated the present study.
\end{acknowledgments}

\appendix

\section{Response-regime bookkeeping}
\label{app:response_matrix}

For completeness, let \(P\) label the probe row and \(\mathcal R\) label the transport regime.  Use
\begin{equation}
        P\in\{V,{\rm th},\eta\},
        \qquad
        \mathcal R\in\{D,{\rm Hyd},{\rm Kin}\}.
\end{equation}
The labels mean voltage, thermal/photothermal, Hall-viscosity-sensitive, diffusive, hydrodynamic, and kinetic, respectively.  A sinusoidal defect produces
\begin{equation}
        \delta M_{P,\mathcal R}^{(N)}(\qv)
        =
        F_N(q)U(\qv)K_P(\qv;B)\chi_{P,\mathcal R}^{(N)}(\qv;\mu,T),
        \label{eq:response_matrix}
\end{equation}
where \(K_P\) is the probe kernel and \(\chi_{P,\mathcal R}\) the response selected by the probe.  To leading order,
\begin{align}
        \chi_{V,\mathcal R}^{(0)}&=-\partial_\Delta\sigma_{\mathcal R}^{(0)},\label{eq:app_chiV}\\
        \chi_{{\rm th},\mathcal R}^{(0)}&=\partial_\Delta\left[\frac{\Delta}{eT}\sigma_{\mathcal R}^{(0)}\right],\label{eq:app_chith}\\
        \chi_{\eta,\mathcal R}^{(2)}&=-\partial_\Delta\sigma_{\mathcal R}^{(2)}.\label{eq:app_chieta}
\end{align}
Here \(\Delta\) is the activation energy to the conducting state, \(\sigma^{(0)}\) the local response, and \(\sigma^{(2)}\) the \(q^2\) nonlocal coefficient.

\section{Real-space multipole check}
\label{app:multipoles}

Let \(\calL_0=-\nabla\cdot\sigma_0\nabla\), with \(\calL_0\phi=s\) for the source field and \(\calL_0^\dagger\psi=w\) for the probe-adjoint field.  A perturbation changes the measurement by
\begin{equation}
        \delta M=-\langle\psi|\delta\calL|\phi\rangle .
\end{equation}
For a small conductivity disk of radius \(a\),
\begin{equation}
        \delta\calL_\sigma\phi=-\nabla\cdot[\delta\sigma\chi_a(\rr)\nabla\phi],
\end{equation}
so
\begin{equation}
        \delta M_\sigma
        =
        -\int\dd^2r\,\delta\sigma\chi_a\nabla\psi\cdot\nabla\phi .
\end{equation}
For source and probe far from the disk this response is dipolar, \(\delta M_\sigma\propto R^{-2}\cos\theta\).  A stress-like perturbation,
\begin{equation}
        \delta\calL_Q\phi=\partial_i\partial_j[Q\chi_a(\rr)\partial_i\partial_j\phi],
\end{equation}
produces
\begin{equation}
        \delta M_Q
        =
        -\int\dd^2r\,Q\chi_a(\partial_i\partial_j\psi)(\partial_i\partial_j\phi),
\end{equation}
which is quadrupolar, \(\delta M_Q\propto R^{-4}\cos2\theta\).  A scalar \(\nabla^4\) perturbation gives \((\nabla^2\psi)(\nabla^2\phi)\) and vanishes in source-free bulk at first order.

\clearpage
\onecolumngrid
\makeatletter
\@removefromreset{equation}{section}
\def\theequation@prefix{S}
\makeatother
\newcommand{\suppsection}[1]{\refstepcounter{section}\section*{S\arabic{section}. #1}}
\setcounter{section}{0}
\setcounter{subsection}{0}
\setcounter{equation}{0}
\setcounter{figure}{0}
\setcounter{table}{0}
\renewcommand{\thesection}{S\arabic{section}}
\renewcommand{\thesubsection}{S\arabic{section}.\arabic{subsection}}
\renewcommand{\theequation}{S\arabic{equation}}
\renewcommand{\thefigure}{S\arabic{figure}}
\renewcommand{\thetable}{S\arabic{table}}
\renewcommand{\theHsection}{supp.S\arabic{section}}
\renewcommand{\theHsubsection}{supp.S\arabic{section}.\arabic{subsection}}
\renewcommand{\theHequation}{supp.S\arabic{equation}}
\renewcommand{\theHfigure}{supp.S\arabic{figure}}
\renewcommand{\theHtable}{supp.S\arabic{table}}

\begin{center}
{\Large\bfseries Supplemental Material: Thermal and viscous contrast in quantum Hall scanning-probe images\par}
\vspace{0.8em}
{\large P.~Shubham~Parashar\par}
\vspace{0.25em}
{\small Independent Researcher, San Diego, California, USA\par}
{\small \texttt{psparash@ucsd.edu}; \texttt{pathaksparash@gmail.com}\par}
\end{center}
\vspace{1em}

\suppsection{Conventions and scope}

The notation follows the main text.  ``Nuisance parameter'' is used in the statistical sense: a fitted parameter that is not the reported target but can overlap with it over the finite measured wave-number band.  The name does not mean that the corresponding physics is unimportant.  After angular averaging the radial measure is $\intq=(2\pi)^{-1}\int_0^\infty q\,\dd q$.

The Landau-level projection of the defect (Sec.~\ref{subsec:projected_vertex}), the thermal triangle and magnetization-current subtraction (Sec.~\ref{subsec:thermal_ratio}), the finite-tip Fisher construction (Sec.~\ref{subsec:fisher_warmup}), the hydrodynamic tensor inversion (Sec.~\ref{subsec:gradient_expansion}), and the Schur-complement algebra (Sec.~\ref{subsec:schur}) are derived in full in the main text and are not repeated here.  This Supplement collects only the material not contained there: the broadened-edge model behind Fig.~\ref{fig:thermal_filter}, two auxiliary figures, the $\beta$-dependence of the Schur error, and the explicit response-regime matrix.

\suppsection{Broadened-edge model for the thermoelectric filter}

Figure~\ref{fig:thermal_filter} replaces the sharp mobility edge of Eq.~\eqref{eq:edge_defs} by a symmetric Gaussian,
\begin{equation}
        \partial_\eps\Phi(\eps)=\frac{\Phi_0}{\sqrt{2\pi}\,w}\exp\!\left[-\frac{(\eps-E_c)^2}{2w^2}\right],
\end{equation}
so that the sharp-edge result $t\,g(t)$ is recovered as $w\to0$.  For this symmetric model the thermoelectric filter $\chi_{\rm th}$ [Eq.~\eqref{eq:chith_def}] remains odd in $t=(E_c-\mu)/T$ and its zero stays pinned at $E_c=\mu$; broadening only lowers the peak and pushes the extrema outward.  The spectral width $w$ is a property of the disorder-broadened edge and is unrelated to the momentum-relaxation coefficient $\Gamma$ of the hydrodynamic tensor---consistent with the two results being set in different regimes (Sec.~\ref{sec:introduction}).

\suppsection{Scalar finite-tip Fisher inverse}

\begin{figure}[tbp]
\centering
\includegraphics[width=0.62\textwidth]{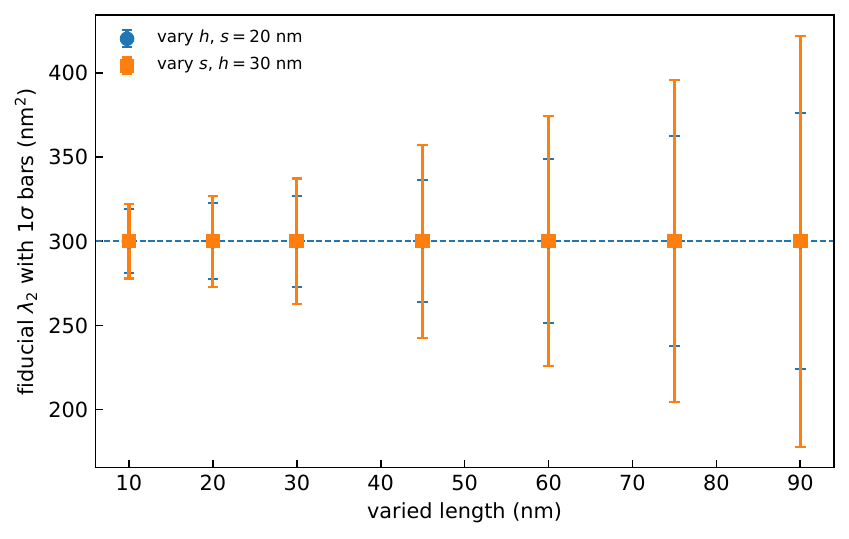}
\caption{Scalar finite-tip inverse.  The markers are placed at a reference value $\lambda_2=300\,\mathrm{nm}^2$ only to display the $1\sigma$ errors.  A simultaneous $a^2q^4$ nuisance parameter is marginalized.}
\label{fig:scalar_inverse_supp}
\end{figure}

The scalar warm-up of Sec.~\ref{subsec:fisher_warmup}---a single target amplitude $\lambda_2$ with a marginalized $a^2q^4$ nuisance $\lambda_4$---gives $\delta\lambda_2=5.410\times10^3\,\mathrm{nm}^2/{\rm SNR}_0$ for the default geometry.  Figure~\ref{fig:scalar_inverse_supp} displays this inverse as a $1\sigma$ error bar; the main text uses the fuller Schur-complement maps of Fig.~\ref{fig:schur_sweeps}.

\suppsection{Field dependence of the Schur error}

Holding the radial weight $W(q;h,s)$ and the nuisance library $\calN$ of Sec.~\ref{subsec:schur} fixed while varying only the tensor vectors $\mathbf v_H,\mathbf v_L,\mathbf v_\perp$ separates two quantities that are easily conflated.  The residual fraction $R_H/(T_H|T_H)_0$ measures template degeneracy after the raw target norm is divided out; the absolute error $\delta d_H$ also carries the raw norm $|\mathbf v_H|^2=\beta^4+6\beta^2+1$.  At the default geometry and ${\rm SNR}_0=200$,
\begin{center}
\begin{tabular}{c c c}
\toprule
$\beta$ & $R_H/(T_H|T_H)_0$ & $\delta d_H$ (nm$^2$)\\
\midrule
0.1 & $1.01\times10^{-2}$ & 187\\
1.0 & $1.01\times10^{-2}$ & 67.8\\
2.0 & $9.91\times10^{-3}$ & 30.3\\
3.0 & $9.89\times10^{-3}$ & 16.6\\
\bottomrule
\end{tabular}
\end{center}
The residual fraction stays near $10^{-2}$ across the range while $\delta d_H$ falls as $|\mathbf v_H|^2$ grows.  This is why the main text calls $\beta=1$ the tensor-orthogonal point rather than the global minimum of the Schur-projected error [Eq.~\eqref{eq:dH_number}]; at larger $\beta$ the practical limit is set by hydrodynamic validity ($|\omega_c\tau_{\rm ee}|\ll1$), heating, and device noise.

\suppsection{Template-overlap matrix}

\begin{figure}[tbp]
\centering
\includegraphics[width=0.58\textwidth]{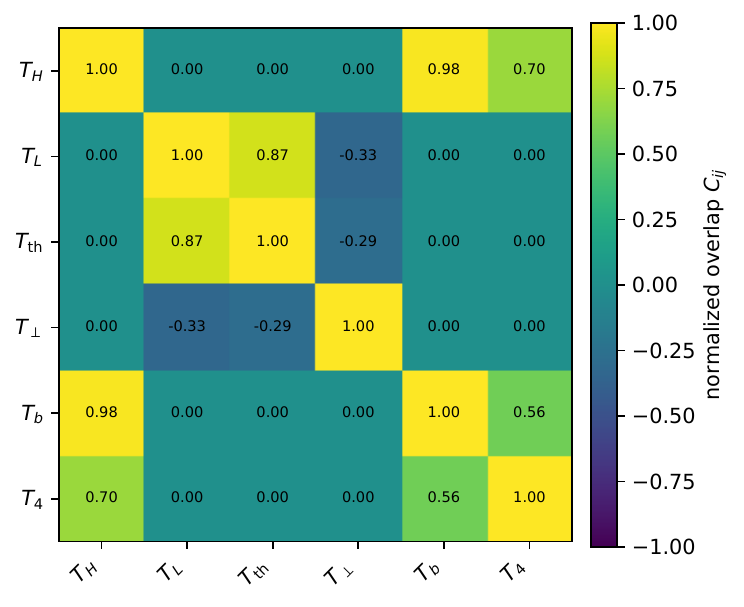}
\caption{Normalized template-overlap matrix for the default geometry.  Values close to $\pm1$ indicate nearly parallel template shapes in the finite measured $q$-band.  The large $T_H$--$T_\zeta$ overlap shows why Hall-odd slip leakage is a serious competing direction for Hall viscometry.}
\label{fig:template_overlap_supp}
\end{figure}

Figure~\ref{fig:template_overlap_supp} shows the normalized template-overlap matrix at the default geometry,
\begin{equation}
        C_{ij}=\frac{(T_i|T_j)_0}{[(T_i|T_i)_0(T_j|T_j)_0]^{1/2}}.
\end{equation}
The large $T_H$--$T_\zeta$ overlap is why the boundary-slip template must be included explicitly: the Schur complement of Eq.~\eqref{eq:schur} subtracts the component of $T_H$ lying in the span of $\calN$, and the leave-one-out test of Sec.~\ref{subsec:schur} identifies $T_\zeta$ as the limiting nuisance.

\suppsection{Explicit response-regime matrix}

The response-regime bookkeeping of Appendix~\ref{app:response_matrix} can be written as a single probe-by-regime matrix,
\begin{equation}
M^{(N)}(\qv)=F_N(q)U(\qv)
\begin{pmatrix}
K_V^D\chi_V^D & K_V^{\rm Hyd}\chi_V^{\rm Hyd} & K_V^{\rm Kin}\chi_V^{\rm Kin}\\
K_{\rm th}^D\chi_{\rm th}^D & K_{\rm th}^{\rm Hyd}\chi_{\rm th}^{\rm Hyd} & K_{\rm th}^{\rm Kin}\chi_{\rm th}^{\rm Kin}\\
K_\eta^D\chi_\eta^D & K_\eta^{\rm Hyd}\chi_\eta^{\rm Hyd} & K_\eta^{\rm Kin}\chi_\eta^{\rm Kin}
\end{pmatrix},
\end{equation}
whose rows are probe channels (voltage $V$, thermal/photothermal ${\rm th}$, Hall-viscosity-sensitive $\eta$) and columns are transport regimes (diffusive $D$, hydrodynamic Hyd, kinetic Kin); the entry $K_P^R\chi_P^R$ suppresses tensor-contraction indices.  The controlled comparisons are
\begin{align}
        \text{fixed probe, changed regime:}\quad&
        M_P^{\rm Hyd}-M_P^D,\quad M_P^{\rm Kin}-M_P^D,\\
        \text{fixed regime, changed probe:}\quad&
        M_{\rm th}^R/M_V^R,\quad M_\eta^R/M_V^R.
\end{align}
Changing both probe and regime in a single step is not a clean diagnostic.  This is the only purpose of the matrix language: it prevents the accidental comparison of unlike filters that the two-regime structure of the paper otherwise invites.

\vspace{1em}
\noindent\textit{The real-space multipole check of Appendix~\ref{app:multipoles} completes the operator picture.  References cited in this Supplement are listed in the bibliography of the main text.}

\end{document}